\documentstyle[preprint,aps]{revtex}
\begin{document}
\topskip 20mm
\draft
\tightenlines
\title{Random Matrices with Correlated Elements: A Model for Disorder with 
Interactions}
\author{Pragya Shukla$^{a,b}$}
\address{(a) Department of Physics, University of Illinois at Chicago, 
Chicago, Illinois-60607, \\
(b) Department of Physics,
Indian Institute of Technology, Kharagpur, India.}
\onecolumn
\date{\today}
\maketitle
\widetext
\begin{abstract}
% .
 	The complicated interactions in presence of disorder lead to a correlated 
randomization of states. The Hamiltonian as a result behaves like a multi-parametric 
random matrix with correlated elements. We show that the eigenvalue correlations of these 
matrices can be described by the single parametric Brownian ensembles. The analogy helps 
us to reveal many important features of the level-statistics in interacting systems e.g. 
 a critical point behavior different from that of non-interacting systems, the possibility 
of extended states even in one dimension and a universal formulation of level correlations.

\end{abstract}

%..
%\begin{multicols}{2}

\section{Introduction}	

	The theoretical formulation of physical properties of electronic systems 
is usually based on independent electron approximation. However, 
experiments on quantum dots as well as extended semi-conducting electron 
heterostructures show that Coulomb interaction between electrons is by no means small 
\cite{us,srp}.
The influence of Coulomb interaction is particularly strong in 
presence of a disordered environment, with  bulk and mesoscopic systems both revealing  
new features \cite{pdkc,alh}. Examples  include Coulomb Blockade phenomena, low 
energy anomalies in transport and thermodynamics coefficients, non-Fermi liquid effects 
in effectively one dimensional structures, various manifestations of the Kondo effects, 
the fractional quantum Hall effects etc. A theory of physical properties including electron 
electron (e-e) interaction is therefore very much required.  

	The presence of interactions along with disorder makes a physical system 
so complex that its properties can be analyzed only by a statistical tool. In past, 
the statistics of single particle states of non-interacting systems with external 
disorder potential has been well-modeled by the Wigner-Dyson ensembles, 
also known as standard Random matrix ensembles (RMT) \cite{meta,gw,alh}. The 
ensembles have also been used successfully to describe the statistical properties of the 
high energy states of many body systems with no disorder e.g. nuclei, atoms etc \cite{meta,gw,alh}. 
The intuition suggests therefore a possibility of random matrix modeling of systems when 
disorder and interactions are both present. Recent studies in this direction have 
led to two types of random matrix models however both impose specific conditions on 
nature and degree of e-e interactions as well as disorder in the system \cite{ajw,js}.  
In this paper, we consider a random matrix model suitable for generic conditions for both 
disorder and interactions and show that the spectral statistics 
can be described by a mathematical formulation analogous to that of non-interacting disordered systems.
 The parameter governing the localization $\rightarrow$ delocalization transition however turns 
out to be different in the two cases. 

%at the critical point of transition is significantly different 
%from the non-interacting case. 

	The complexity of many body interactions or external disorder potential 
associates a degree of uncertainty in the exact determination of the Hamiltonian.   
As a result, some (or all) elements of the Hamiltonian matrix behave like random 
variables. The distribution properties of the matrix elements are governed by the 
nature of interactions and disorder; their distributions  need not be same, may or may not 
be correlated and some of them can be non-random too. For example, a single particle 
Hamiltonian with a "white noise" disorder potential (that is, no e-e interaction 
or impurity interaction) can be modeled by a random  matrix with uncorrelated elements. 
However the presence of local interactions among impurities or electrons or both will result 
in a correlated randomization of matrix elements of the Hamiltonian.    
Unfortunately not much information for random matrices with correlated elements  
has been available so far. Our objective in this study is to suggest a 
way to fill in this information gap. Here we show that the eigenvalue distributions
of various ensembles, with correlated matrix elements and a
multi-parametric probability density,  appear as non-equilibrium stages of 
a Brownian type diffusion process. The diffusing eigenvalues evolve with 
respect to a single parameter which is a function of the correlation 
strengths between various matrix elements. The parameter is therefore 
related to  complexity of the system represented by the ensemble and 
can be termed as  "complexity" parameter. The solution of the diffusion 
equation for a given value of complexity parameter gives, therefore, the
distribution of eigenvalues, and thereby their correlations, for 
corresponding system. As discussed later, the diffusion equation can be solved 
by using its analogy with the one governing the evolution of Brownian ensembles \cite{ps1}. 
The analogy also helps in theoretical formulation of the level-density correlations 
of interacting disordered systems.

        The paper is organized as follows. In section II, we derive the
diffusion equation for the matrix elements of the Hermitian operators governing the 
dynamics in complex systems. For a clear exposition of our technique, we first 
consider the cases with $2^{nd}$ and $3^{rd}$ order matrix elements correlations and 
generalize to $n^{th}$ order correlations later on. The diffusion equation for the matrix 
elements  is then used to obtain the equation governing the evolution of the eigenvalues 
in section III. To maintain the flow of the discussion, only relevant steps are given 
in the sections II, III; the details can be found in the appendices. 
As mentioned above, the evolution equation for the 
eigenvalues turns out to be a very well-known equation and calculation of the eigenvalue 
correlations from the equation  has been discussed in detail many times. We therefore 
avoid the repetition but give a brief discussion to keep the article self-content.  
This is followed, in section IV, by a discussion of the application of our technique to 
two well-known systems. We conclude in section V by summarizing our main results.

\section{Single Parametric Evolution of Matrix Elements}

	Let us consider a complex system represented by a $N\times N$ Hermitian 
matrix $H$ with $H_{kl}=\sum_{s=1}^\beta (i)^{s-1} H_{kl;s}$ as its matrix-elements.
The subscript $s$ to a variable refers to one of its components with $\beta$ 
as their total number. The parameter $\beta$ contains the information about underlying 
symmetry of the system. For example, for systems with time-reversal symmetry and 
integer angular momentum, the Hamiltonian in a generic representation is a 
real-symmetric matrix which gives $\beta=1$. The Hamiltonians for system without 
time reversal symmetry are, in general, complex Hermitian which gives  $\beta=2$. 
Due to its Hermitian nature, the independent real parameters  
$H_{kl;s}$, which determine all matrix-elements of $H$, are ${\tilde M}=N+N(N-1)\beta/2$ 
in number. For notational simplification, let us denote them by $H_{\mu}$ where 
$\mu\equiv \{kl;s\}$ is a single index which can take a value from $1\rightarrow M$.

 	The elements of $H$ describe the overlap between various  states of the 
basis in which $H$ is represented. A complicated nature of interactions or presence 
of disorder in the system associates a deterministic uncertainty with the matrix elements 
$H_{\mu}$. As the interaction between various states is governed 
by the nature and complexity of the region, the randomness associated with  the elements 
$H_{\mu}$ can be of various types. In general, the complexity of a region 
can cause multiple interactions between basis states, resulting in correlations between matrix 
elements. In this section, we consider the cases which can be modeled by an ensemble 
described by a probability density ${\tilde\rho(H,b)} \propto {\rm e}^{-F(H)}$ with 
function $F(H)$ as a sum over various combinations of the matrix elements of $H$.

%For clarification of the ideas, we first discuss the cases with second and third order 
%correlations between matrix elements which is later on generalized to the case with 
%the $n^{\rm th}$ order correlations.    

%Using maximum entropy principle subjected to known correlations 
%$<M_{ij}M_{kl}>$, such cases can  be modeled by an ensemble described by a probability 
%density ${\tilde\rho(M,b)}=C {\rho(M,b)}$ 

%Here we use notations which can easily be generalized for the higher order 
%correlation between matrix elements.

\subsection{Correlated Gaussian Case}

	A Gaussian ensemble of Hermitian matrices with correlated 
elements can be described by  a matrix elements distribution 
\begin{eqnarray}
\tilde{\rho} (H,b)=C{\rm exp}[ - \sum_{\mu_1=1}^M b_{\mu_1} H_{\mu_1}
-\sum_{\mu_1,\mu_2=1}^M  b_{\mu_1 \mu_2}
 H_{\mu_1} H_{\mu_2})] = C \rho(H,b)
\end{eqnarray}
with $C$ as a constant and $b$ as the set of coefficients $b_{\mu_1}$ and 
$b_{\mu_1 \mu_2}$. Here the subscripts to a coefficient are indicators of the 
terms present in the product of which it is a coefficient.   
Further, in $\sum_{\mu_1,\mu_2}$, similar pairs are included only once. 

The distribution parameters $b_{2;\mu_1,\mu_2}$ are the measures of correlations between
pairs of the matrix elements: 
$<H_{\mu_1} H_{\mu_2}> = {\partial {\rm log}C\over \partial b_{2;\mu_1\mu_2}}$. 
In general, different system conditions can give rise to 
different sets of distribution parameters $b$. A slight perturbation of the system 
due to a change in its parameters perturbs the matrix elements 
and therefore the probability density $\rho(H,b)$.
In the following, we consider a particular flow in the matrix space 
generated by an operator $L$ (describing the diffusion of the matrix elements
with a constant drift and confined by a quadratic potential),
\begin{eqnarray}
L = \sum_{\mu}{\partial \over \partial H_{\mu}}\left[
{g_{\mu}\over 2}
 {\partial \over \partial H_{\mu}} + \gamma H_{\mu}\;\right]
\end{eqnarray}
where $g_{\mu}=1+\delta_{\mu}$ with $\delta_{\mu}=1$ for $\mu=(kk;s)$ and 
$\delta_{\mu}=0$ for $\mu=(kl;s)$, $k\not=l$.
 The parameter $\gamma$ is arbitrary, giving the freedom to choose the 
end of the evolution\cite{ps1}. The choice of the above form of $L$ is 
motivated by our desire to obtain the equation governing the eigenvalue-dynamics 
in a well-known mathematical form. 

	The evolution of the probability density $\rho(H,b)$, generated by operator $L$ 
in the matrix space, is related to a multi-parametric flow in $b$-space. This can be 
shown as follows. The probability density being a function of both $H$ and $b$, 
the rates of change of $\rho$ with variation of matrix elements and parameters $b$
can be given as 
%
%the derivatives of $\rho$ with respect to matrix elements can be expressed in 
%terms of the parametric derivatives,
%
\begin{eqnarray}
{\partial \rho \over \partial H_{\mu_1}} &=& 
 -\left[ b_{\mu_1} + \sum_{\mu_2} \eta^{(\mu_1)}_{\mu_1\mu_2} 
b_{\mu_1 \mu_2}  H_{\mu_2} \right] \rho \\
{\partial \rho \over\partial b_{\mu_1\mu_2}} 
&=&  - H_{\mu_1} H_{\mu_2}\rho \\
{\partial \rho\over \partial b_{\mu_1}} &=& -H_{\mu_1} \rho 
\end{eqnarray}
%
%with $G_{ijkl}= 2$ for $\{ij\}\not\equiv \{kl\}, $G_{ijkl}= 1$ for $\{ij\}\equiv \{kl\}$. 
%
where $\eta^{(\mu_1)}_{\mu_1\mu_2}$ is the frequency of occurrence of the term $H_{\mu_1}$ in the 
 combination $H_{\mu_1} H_{\mu_2}$ (i.e  $\eta^{(\mu_1)}_{\mu_1\mu_2}=2$ for $\mu_1=\mu_2$ 
and 1 for $\mu_1 \not=\mu_2$.  
With the help of eq.(3) and (4), a drift in the matrix space can be written in terms of a 
drift in parametric space, 
\begin{eqnarray}
\sum_{\mu_1} H_{\mu_1} {\partial \rho\over \partial H_{\mu_1}}
&=& - \left[ \sum_{\mu_1} c_{\mu_1} H_{\mu_1}\rho + 
\sum_{\mu_1,\mu_2}  c_{\mu_1\mu_2} H_{\mu_2} H_{\mu_1}\right] \rho \nonumber \\
&=& \sum_{\mu_1} c_{\mu_1} {\partial \rho\over \partial b_{\mu_1}}
+\sum_{\mu_1,\mu_2}  c_{\mu_1\mu_2} 
{\partial \rho\over \partial b_{\mu_1\mu_2}}
\end{eqnarray}
where $c_{1;\mu}= b_{\mu}$ and 
$c_{\mu_1\mu_2}=\eta^{(\mu_1)}_{\mu_1\mu_2} b_{\mu_1\mu_2}$. 
Similarly a diffusion in the matrix space can be expressed as a combination of 
drifts in parametric space, 
\begin{eqnarray}
{\partial^2 \rho \over \partial H_{\mu_1}^2} &=& (c_{\mu_1}^2- c_{\mu_1\mu_1}) \rho + 
2\sum_{\mu_2} c_{\mu_1} c_{\mu_1\mu_2} H_{\mu_2} + 
\sum_{\mu_2,\mu_3} c_{\mu_1\mu_2} c_{\mu_1\mu_3} H_{\mu_2} H_{\mu_3} \rho  \\
&=& (c_{\mu_1}^2-c_{\mu_1\mu_1})\rho -2\sum_{\mu_2} c_{\mu_1}c_{\mu_1\mu_2}
 {\partial \rho\over \partial b_{\mu_2}}
 -\sum_{\mu_2,\mu_3} c_{\mu_1\mu_3} c_{\mu_2\mu_3}
{\partial \rho\over \partial b_{\mu_2\mu_3}}
\end{eqnarray}

 A substitution of above equalities in eq.(2) gives us the following,
\begin{eqnarray}
 L {\rho} = T {\rho} + {\tilde C} \rho 
\end{eqnarray}
with $\tilde C=\sum_{\mu_1}[\gamma+ {g_{\mu_1}\over 2}(c^2_{\mu_1}-c_{\mu_1\mu_2})]$ 
and $T$ as the generator of the dynamics in the parametric space,
\begin{eqnarray}
T \equiv \sum_{\mu_1} f_{\mu_1}{\partial  \over \partial b_{\mu_1}}
+ \sum_{\mu_1,\mu_2} f_{\mu_1\mu_2}{\partial  \over \partial b_{\mu_1\mu_2}}.
\end{eqnarray}
Here $f_{\mu_1}=\gamma c_{\mu_1}-\sum_{\mu_2}g_{\mu_2} c_{\mu_2} c_{\mu_1\mu_2}$ and  
$f_{\mu_1\mu_2}=\gamma c_{\mu_1\mu_2} - 
(1/2)\sum_{\mu_3} g_{\mu_3} c_{\mu_1\mu_3} c_{\mu_2\mu_3}$.

The above equation appears complicated, with many parametric derivatives present 
on its right side. However it is possible to map the multi-parametric flow
in the $M$-dimensional $b$-space to a single parametric drift in another
parametric space, say $y$-space, consisting of variables $y_i$,
$i=1 \rightarrow M$ where $M=[3+{\tilde M}]{\tilde M}/2$. In other words, the 
generator $T$ of the flow in the $b$-space can be reduced to a partial derivative with
respect to just one $y$-space variable, say $y_1$:
\begin{eqnarray}
T(y[b]) \rho \equiv {\partial \rho \over\partial y_1} |_{y_2,..,y_M}
\end{eqnarray}

The desired transformation $b \rightarrow y$ required to convert eq.(10) into the form (11) 
can be obtained as follows. 
By using the definition
${\partial \over \partial x}=\sum_{k=1}^M {\partial y_k \over \partial x}
{\partial \over \partial y_k}$, with $x$ as various $b$ parameters,  
% with $D_{kj}\equiv {\partial y_k \over \partial b_{p(r)}}$, 
$T(b)$ (eq.(10)) can
 be transformed in terms of the derivatives with respect to $y$,
\begin{eqnarray}
T \rho = \sum_k A_k {\partial \rho\over \partial y_k}
\end{eqnarray}
where 
\begin{eqnarray}
A_k \equiv \sum_{\mu_1} f_{\mu_1}{\partial y_k \over \partial b_{\mu_1}}
+ \sum_{\mu_1,\mu_2} f_{\mu_1\mu_2}{\partial  y_k\over \partial b_{\mu_1\mu_2}}.
\end{eqnarray}
The eq.(12) can be reduced in the desired form of eq.(11), if the transformation 
$b\rightarrow y$ satisfies following condition:
\begin{eqnarray}
A_k = \delta_{k1} \qquad\qquad {\rm for}\; k=1\rightarrow M
\end{eqnarray}
The parameters $y$ as a function of $b$ can now be obtained by solving the set of conditions (14), 

\begin{eqnarray}
y_1 &=& \sum_{\mu_1} \int {\rm d}b_{\mu_1} z^{(1)}_{\mu_1} G 
+ \sum_{\mu_1,\mu_2} \int {\rm d}b_{\mu_1 \mu_2} z^{(1)}_{\mu_1\mu_2} G 
+ {\rm constant }
\end{eqnarray}
where $G=[\sum_{\mu_1} z^{(1)}_{\mu_1} f_{\mu_1} + 
\sum_{\mu_1,\mu_2} z^{(1)}_{\mu_1\mu_2} f_{\mu_1\mu_2}]^{-1}$ and the set $z^{(1)}$ of 
the functions $z^{(1)}_{\mu_1}, z^{(1)}_{\mu_1\mu_2}$ are 
chosen such that the ratio 
\begin{eqnarray}
\left[\sum_{\mu_1} z^{(1)}_{\mu_1} {\rm d}b_{\mu_1} +
\sum_{\mu_1\mu_2} z^{(1)}_{\mu_1\mu_2} {\rm d}b_{\mu_1\mu_2} \right] G
\end{eqnarray} 
is a complete differential (see appendix B). Note it is possible  that the ratio (16) 
can be made an exact differential for different sets $z^{(1)}$ which will lead to different 
 solutions of $y$. However any two of such solutions for $y$ are different from each 
 other only by a constant (appendix B). 

The  conditions (14) further imply that parameters $y_k$, $k>1$ behave as the constants of 
the dynamics (generated by $L$) in the matrix space, 
\begin{eqnarray}
y_{k} &=&  \sum_{\mu_1} \int {\rm d}b_{\mu_1} z^{(k)}_{\mu_1}  
+ \sum_{\mu_1,\mu_2} \int {\rm d}b_{\mu_1 \mu_2} z^{(k)}_{\mu_1\mu_2}  
+ {\rm constant} \qquad {\rm for}\; k>1
\end{eqnarray}
with 
$\sum_{\mu_1}  z^{(k)}_{\mu_1} f_{\mu_1}  + 
\sum_{\mu_1,\mu_2}  z^{(k)}_{\mu_1\mu_2} f_{\mu_1\mu_2}=0$ for $k>1$.

The substitution of  eq.(11) in eq.(9),  gives the single parametric 
evolution of the joint probability density $\rho(H)$  in the matrix space

\begin{eqnarray}
\sum_{\mu}{\partial \over \partial H_{\mu}}\left[
{\partial {\tilde\rho}\over \partial H_{\mu}} + \gamma H_{\mu}\; {\tilde\rho}\right]
&=& {\partial {\tilde\rho} \over\partial y_1}
\end{eqnarray}
where $y_1$ is given by eq.(15). As the distribution parameters depend 
on the complexity of the system, $y_1$ can be termed as the complexity parameter.

	The parametric space transformation $b \rightarrow y$ maps the 
probability density $\rho (H,b)$ to $\rho(H,y(b))$. As a result, $\rho$ depends on  
various parameters $y_k$, $k=1 \rightarrow {\tilde N}$. However eq.(18) implies that 
the diffusion, generated by the operator $L$ in the matrix space, is governed by $y_1$ only; 
the rest of them, namely $y_k$, $k>1$, remain constant during the evolution. Note it is always 
possible to define a transformation from the set $b \rightarrow y$ with $y_k$, $k>1$ as 
 constants of the dynamics generated by $L$. This can be explained as follows. 
A matrix element, say $H_{ij}$, describes how a basis state $\psi_i$ interacts with state 
$\psi_j$ through $H$. This results in dependence of the matrix element correlations 
and, thereby, of the set $b$, on the basis parameters e.g. basis indices.
As the basis remains fixed during the evolution, the suitable 
functions of basis parameters can be chosen to play the role of $y_k$, $k>1$.
(Note a similar transformation has been used to obtain a single parametric evolution of 
multi-parametric Gaussian ensembles of Hermitian matrices; see \cite{ps1} for details).

\subsection{Non-Gaussian Case with third order matrix elements correlations}

	Let us consider an ensemble of Hermitian matrices with a $3^{\rm rd}$ correlations 
among its matrix elements, described by a  probability density 
\begin{eqnarray}
{\tilde{\rho}} (H,b)=C{\rm exp}{[ - \sum_{\mu_1=1}^M  b_{\mu_1} H_{\mu_1} 
-\sum_{\mu_1,\mu_2=1}^M  b_{\mu_1 \mu_2} H_{\mu_1} H_{\mu_2} 
-\sum_{\mu_1,\mu_2,\mu_3=1}^M b_{\mu_1 \mu_2 \mu_3} 
H_{\mu_1} H_{\mu_2} H_{\mu_3}]} 
= C \rho(H,b)
\end{eqnarray}

Again the parameters $b_{\mu_1\mu_2\mu_3}$ are the measures of 
correlations among  three matrix elements: 
$<H_{\mu_1} H_{\mu_2} H_{\mu_3}> = 
{\partial {\rm log}C\over \partial b_{\mu_1\mu_2\mu_3}}$. 
Proceeding as in the Gaussian case, we get 

\begin{eqnarray}
{\partial \rho \over \partial H_{\mu_1}} &=& 
-\left[ c_{\mu_1} + \sum_{\mu_2}  c_{\mu_1 \mu_2}  H_{\mu_2} 
+\sum_{\mu_2,\mu_3}  c_{\mu_1 \mu_2 \mu_3} H_{\mu_2} H_{\mu_3}  \right] \rho \\
{\partial \rho \over\partial b_{\mu_1\mu_2 \mu_3}} 
&=&  - H_{\mu_1} H_{\mu_2} H_{\mu_3}\rho 
\end{eqnarray}
%
%with $G_{ijkl}= 2$ for $\{ij\}\not\equiv \{kl\}, $G_{ijkl}= 1$ for $\{ij\}\equiv \{kl\}$. 
The derivatives ${\partial \rho \over\partial b_{\mu}}$ and 
${\partial \rho \over\partial b_{\mu_1\mu_2}}$ remain same as in the 
Gaussian case (given by eqs.(4,5)),

\begin{eqnarray}
\sum_{\mu_1} H_{\mu_1} {\partial \rho\over \partial H_{\mu_1}}
&=& \sum_{\mu_1} c_{\mu_1} {\partial \rho\over \partial b_{\mu_1}}
+\sum_{\mu_1,\mu_2} c_{\mu_1\mu_2} 
{\partial \rho\over \partial b_{\mu_1\mu_2}}
+\sum_{\mu_1,\mu_2,\mu_3} c_{\mu_1\mu_2\mu_3} 
{\partial \rho\over \partial b_{\mu_1\mu_2\mu_3}}
\end{eqnarray}

where $c_{\mu_1\mu_2\mu_3}=\eta^{(\mu_1)}_{\mu_1\mu_2\mu_3} b_{\mu_1\mu_2\mu_3}$ with 
$\eta^{(\mu_1)}_{\mu_1\mu_2\mu_3}$ as the frequency of occurrence of the term $H_{\mu_1}$ 
in the combination $H_{\mu_1} H_{\mu_2} H_{\mu_3}$. The second order derivative of $\rho$ 
with respect to $H_{\mu}$ can  be calculated by differentiating eq.(20) which, on 
combining with eq.(21), again leads to the form $L{\tilde \rho} = (T + {\tilde C})\rho$. 
Here $L$ is still given by eq.(2) 
%
%\begin{eqnarray}
%L\rho = T \rho + S \rho
%\end{eqnarray}
however the generator $T$ now contains the first order parametric derivatives 
as well as their products,

\begin{eqnarray}
 T \rho &=&\sum_{\mu_1} f_{\mu_1}{\partial\rho \over \partial b_{\mu_1}}
+ \sum_{\mu_1,\mu_2} f_{\mu_1\mu_2}{\partial  \rho\over \partial b_{\mu_1\mu_2}}
+ \sum_{\mu_1,\mu_2,\mu_3} f_{\mu_1\mu_2\mu_3} 
{\partial  \rho \over \partial b_{\mu_1\mu_2\mu_3}}  \nonumber \\
 &+& \sum_{\mu_1,\mu_2,\mu_3 \mu_4 \mu_5} c_{\mu_1\mu_2\mu_3} c_{\mu_1\mu_4\mu_5} 
{1\over \rho} {\partial \rho \over \partial b_{\mu_2\mu_3}}
{\partial \rho \over \partial b_{\mu_4\mu_5}}
\end{eqnarray}
with
$f_{\mu_1}=\gamma c_{\mu_1} - (1/2)\sum_{\mu_2} g_{\mu_2} [2 c_{\mu_2}c_{\mu_1\mu_2}+
 c_{\mu_2\mu_2\mu_1}(\eta^{(\mu_1)}_{\mu_2\mu_2\mu_1}-1)]$,  
$f_{\mu_1\mu_2}=\gamma c_{\mu_1\mu_2} - (1/2)
\sum_{\mu_3} g_{\mu_3} [c_{\mu_1\mu_3} c_{\mu_2\mu_3} + 2 c_{\mu_3} c_{\mu_1\mu_2\mu_3}]$ 
and $f_{\mu_1\mu_2\mu_3}=\gamma c_{\mu_1\mu_2\mu_3} - 
  \sum_{\mu_4} g_{\mu_4} c_{\mu_2\mu_4} c_{\mu_1\mu_3\mu_4}$. 

 Similar to Gaussian case, eq.(23) can also be reduced to a single parametric derivative, 
namely form (11), by using a transformation from the set $b$ to another set $y$. 
The operator $T$ in this case transforms  as  
\begin{eqnarray}
T \rho =  \sum_k A_k {\partial \rho\over \partial y_k}
+\sum_{i,k} B_{ik} {\partial \rho\over \partial y_i }
 {\partial {\rm log}\rho\over \partial y_k }
\end{eqnarray}
where 
\begin{eqnarray}
A_{k}&=&\sum_{\mu_1} f_{\mu_1}{\partial y_k\over \partial b_{\mu_1}}
+ \sum_{\mu_1,\mu_2} f_{\mu_1\mu_2}{\partial y_k \over \partial b_{\mu_1\mu_2}}
+ \sum_{\mu_1,\mu_2,\mu_3} f_{\mu_1\mu_2\mu_3} 
{\partial y_k \over \partial b_{\mu_1\mu_2\mu_3}}  \\
B_{ik } &=& \sum_{\mu_1,\mu_2,\mu_3 \mu_4 \mu_5} c_{\mu_1\mu_2\mu_3} c_{\mu_1\mu_4\mu_5} 
{\partial y_i \over \partial b_{\mu_2\mu_3}}
{\partial y_k \over \partial b_{\mu_4\mu_5}}
\end{eqnarray}
The reduction of eq.(24) for $T\rho$ to a single derivative ${\partial\rho\over \partial y_1}$ 
will impose following conditions on the transformation $b \rightarrow y$:
\begin{eqnarray} 
A_k &=& N \delta_{k1} \\ 
 B_{ik}+B_{ki} &=& 0   
\end{eqnarray}
By solving  conditions (27), $y_1$ can be obtained as a function of the parameters $b$,

\begin{eqnarray}
y_1= \sum_{\mu_1} \int {\rm d}b_{\mu_1} z^{(1)}_{\mu_1} G 
+ \sum_{\mu_1,\mu_2} \int {\rm d}b_{\mu_1 \mu_2} z^{(1)}_{\mu_1\mu_2} G
+ \sum_{\mu_1,\mu_2,\mu_3} \int {\rm d}b_{\mu_1 \mu_2 \mu_3} z^{(1)}_{\mu_1\mu_2\mu_3} G
+constant
\end{eqnarray}
where 
$G=[\sum_{\mu_1} z^{(1)}_{\mu_1} f_{\mu_1}+\sum_{\mu_1,\mu_2} z^{(1)}_{\mu_1\mu_2} f_{\mu_1\mu_2}
 +\sum_{\mu_1\mu_2\mu_3} z^{(1)}_{\mu_1\mu_2\mu_3} f_{\mu_1\mu_2\mu_3}]^{-1} $. 
Again the choice of the functions $z^{(1)}$ is so as to make the ratio
\begin{eqnarray}
\left[\sum_{\mu_1} z^{(1)}_{\mu_1} {\rm d}b_{\mu_1} +
\sum_{\mu_1\mu_2} z^{(1)}_{\mu_1\mu_2} {\rm d}b_{\mu_1\mu_2} 
+\sum_{\mu_1\mu_2\mu_3} z^{(1)}_{\mu_1\mu_2\mu_3} {\rm d}b_{\mu_1\mu_2\mu_3}\right] G 
\end{eqnarray} 
a complete differential. However the set $z^{(1)}$ in this case have to satisfy 
an extra set of conditions given by eq.(29). 
Again as in the Gaussian case, the parameters $y_j$, $j>1$ 
behave as constants of dynamics for this case too, 
\begin{eqnarray}
y_{k} &=&  \sum_{\mu_1} \int {\rm d}b_{\mu_1} z^{(k)}_{\mu_1}
+ \sum_{\mu_1,\mu_2} \int {\rm d}b_{\mu_1 \mu_2} z^{(k)}_{\mu_1\mu_2}  
+ \sum_{\mu_1,\mu_2\mu_3} \int {\rm d}b_{\mu_1 \mu_2} z^{(k)}_{\mu_1\mu_2\mu_3}  \qquad {\rm for}\; k>1
\end{eqnarray}
with arbitrarily chosen functions $z^{(k)}_{\mu}$ satisfy the constraint 
$\sum_{\mu_1}  z^{(k)}_{\mu_1} f_{\mu_1}  +
\sum_{\mu_1,\mu_2}  z^{(k)}_{\mu_1\mu_2} f_{\mu_1\mu_2}
+\sum_{\mu_1,\mu_2}  z^{(k)}_{\mu_1\mu_2\mu_3} f_{\mu_1\mu_2\mu_3}=0$ for $k>1$  as well as
the conditions given by eq.(28).
                              
	Using the $b\rightarrow y$ transformation given by eq.(29,31), the dynamics of $\rho(H)$ with 
third order matrix element correlations  can again be described by eq.(18). Note although 
the evolution is governed by same equation, however the complexity parameters are different for 
Gaussian and non-Gaussian cases.

\subsection{General Case}

	A  generalized ensemble of Hermitian matrices with correlated 
elements can be described by  a matrix elements distribution 
${\tilde\rho(H)}=C {\rho(H)}$ where 
\begin{eqnarray}
\rho(H) = \prod_{r=1}^n {\rm exp}\left[-
\sum_{p(r)} b_{p(r)} \left({\prod}_{j=1}^r H_{\mu_{P_j }}\right)\right]
\end{eqnarray}
with $C$ as a normalization constant. 
Here symbol $p(r)$ refers to a combination of $r$ elements chosen 
from a set of total ${\tilde M}=N+N(N-1)\beta/2$ elements of upper (or lower) diagonal matrix; 
note the terms present in a
given combination need not be all different. The $\prod_{j=1}^r H_{\mu_{P_j}}$ 
implies a product over $r$ terms present in the $p^{\rm th}$ combination with 
 coefficient $b_{p(r)}$ as a measure of their correlation: 
$<{\prod^r}_{j=1} H_{\mu_{p_j}}>= {\partial {\rm log} C\over \partial b_{p(r)}}$. 
The $\sum_{p(r)}$ is a sum over all possible combinations (total $(\tilde M)^r$) of 
$r$ elements chosen from a set of total ${\tilde M}$ of them; the sum includes only 
different combinations.   
(Henceforth subscript $p(r)$ will be written as $p$ only unless
 details required for clarification).

Using eq.(32), the partial derivatives of $\rho$ in matrix space and parametric 
space can be given as
\begin{eqnarray}
{\partial \rho \over \partial H_{\mu}} &=& -
\sum_{r=1}^n \sum_p \left[ b_p 
\left(\prod_{j=1}^r H_{\mu_{p_j}} \right)\sum_{j=1}^r
{\partial {\rm log} H_{\mu_{p_j}}\over \partial H_{\mu}}\right]\;\rho \\
{\partial \rho \over \partial b_{p(r)}} &=& - \left({\prod}_{j=1 }^r H_{\mu_{P_j }} \right)\; \rho 
\end{eqnarray}
As a result, a drift in the matrix space can be related to a drift in the 
parametric space: 
\begin{eqnarray}
\sum_{\mu} H_{\mu} {\partial \rho\over \partial H_{\mu}}
= \sum_{\mu} \sum_{r,p(r)} \eta^{(\mu)}_p b_p {\partial \rho\over \partial b_p}
\end{eqnarray}

where the term $\eta^{(\mu)}_{p(r)}=\sum_{m=1}^r \delta(\mu_{p_m};\mu)$ counts the frequency 
of occurrence of the element $H_{\mu}$ in the $p^{th}$ combination of
$r$ elements ($\delta(\mu_{p_j},\mu)=1$ if $\mu_{p_j}=\mu$ and $0$ is 
$\mu_{p_j}\not=\mu$.
Similarly a diffusion in the matrix space is related to a nonlinear flow 
in the parametric space,

\begin{eqnarray}
{\partial^2 \rho \over \partial H_{\mu}^2} 
&=&\sum_{r=1}^n \sum_{p(r)} \eta^{(\mu)}_p b_p \left[ 
\sum_{r'>(n+2-r)} \sum_{p'(r')} \eta^{(\mu)}_{p'} b_{p'} 
{\partial\rho\over \partial b_{p+\mu}}{\partial\rho\over \partial b_{p'+\mu}}\; 
\rho^{-1}  - \sum_{r'\le (n+2-r)} \sum_{p'(r')} \eta_{p';\mu} b_{p'}
{\partial\rho\over \partial b_{p'+p-\mu^2}}
+ (\eta^{(\mu)}_p - 1) {\partial \rho\over \partial b_{p-\mu^2} }\right]
\end{eqnarray}
Here the notation $A+B$ 
refers to a combination which contains the elements of both $A$ and $B$. 
Similarly $A-B$ indicates dropping of all  
the elements of $B$ from $A$. Further the notation $\mu^k$ is used to denote 
a combination $H_{\mu}^k$ (that is, $k^{\rm th}$ power of $H_{\mu}$).     

% where $u$ refers to a combination of $r''=r+r'-\mu^2$ elements, 
% such that $u(r'') \equiv p(r)+p'(r')-\mu^2$. 

Again, the matrix space flow generated by the operator $L$ (eq.(2)) can be related to 
a parametric flow generated by the operator $T$, 
$L \rho = T \rho + {\tilde C} \rho $, 
where $T$ is now given by 
\begin{eqnarray}
 T \rho \equiv \sum_{\mu} {g_{\mu}\over 2} \sum_{r=1}^n \sum_{p(r)}
\left[ \sum_{r'>(n+2-r)}^m \sum_{p'(r')}  h_{pp'}  
 {\partial {\rm log} \rho\over \partial b_{p'} }
 + f_p \right]{\partial \rho\over \partial b_p }
\end{eqnarray}

with $h_{pp'}=b_u b_{u'} [\sum_{\mu} g_{\mu} \eta^{(\mu)}_u \eta^{(\mu)}_{u'}]$
 and $f_p=-\sum_{r',p'(r')}b_{p'} b_q
[\sum_{\mu} g_{\mu} \eta^{(\mu)}_{p'} \eta^{(\mu)}_q] + [\sum_{\mu} 
\eta^{(\mu)}_v g_{\mu} (\eta^{(\mu)}_v-1)] b_{v} 
+2 \gamma [\sum_{\mu} \eta^{(\mu)}_{p}] b_p $.
%
% with the combination $q \equiv p'(r')-p(r)+(kl)^2$ and $v=p(r)+(kl)^2$.
here $q$ refers to a combination of $r_1=r'-r+2$ elements, 
such that $q(r_1) \equiv p'(r')-p(r)+(kl)^2$. Similarly  $u,u',v$ refer to combinations 
$u(r_0)=p(r)+1$, $u(r_1)=p'(r')+1$ and $v(r_2)=p(r)+2$ of $r_0=r+1$, 
$r_1=r+1$ and $r_2=r+2$ elements respectively. 
%The $+$ sign between two combinations $A$ and $B$ 
%results in a combination which contains the elements of both $A$ and $B$. 
%Similarly the $-$ sign indicates dropping of all the elements of $B$ from $A$. 
%Further the notation $(kl)^m$ is used to denote a combination $H_{kl;s}^m$ 
%(that is, $m^{\rm th}$ power of $H_{kl;s}$).    

%The further analysis of eq.(4) and its application to obtain the 
%eigenvalue correlations depends on nature of the operator $T$, a 
%function of $M$ variables $b_{p(r)}$ where $M=\sum_{r=1}^n {\tilde M}^r$. 

%As shown below, it is possible to map the multi-parametric flow 
%in the $M$-dimensional $b$-space to a single parametric drift in another  
%parametric space, say $y$-space, consisting of variables $y_i$, 
%$i=1 \rightarrow M$. In other words, the generator $T$ of the 
%flow in the $b$-space can be reduced to a partial derivative with 
%respect to just one $y$-space variable, say $y_1$:  

The desired transformation $b \rightarrow y$ required to convert  
eq.(37) into form $T(y[b]) \rho \equiv {\partial \rho \over\partial y_1}$
can be obtained as follows. The substitution of 
${\partial \over \partial b_{p(r)}}=\sum_{k=1}^M D_{kj} 
{\partial \over \partial y_k}$ in eq.(37) with 
$D_{kj}\equiv {\partial y_k \over \partial b_{p(r)}}$ transforms $T$ 
in the form of eq.(24) where $A_k$ and $B_{ik}$ are now given as
\begin{eqnarray}
A_k (y,b) = \sum_r \sum_{p(r)} f_{p} D_{kp} \\
B_{ik}(y,b) =\sum_{r,p(r)}\sum_{r',p'(r')} h_{pp'}(x) D_{ip} D_{kp'} 
\end{eqnarray}
Again, for desired reduction of $T$ to ${\partial \rho \over\partial y_1}$, 
the transformation $b\rightarrow y$ should satisfy conditions given by eq.(27,28).
The conditions (27) can then be solved to obtain the variables in set $y$ as  functions 
of variable in set  $b$ (see appendix B) 
%\cite{sne} for the methods to solve such equations): 
\begin{eqnarray}
y_k= \sum_{r,p(r)} \int  {\rm d} b_p z^{(k)}_p [1-\delta_{k1}+G_k] + {\rm constant}
\end{eqnarray}
with 
$G_k= \delta_{k1} \left[\sum_{r,p(r)} z^{(1)}_{p} f_{p}\right]^{-1} $ and 
$z^{(k)}_p$ as arbitrary functions
which make  $\sum_{r,p(r)} {\rm d} b_p z^{(k)}_p [1+ (G-1)\delta_{k1}]$  
an exact differential and satisfy the constraint 
$\sum_{r,r'}\sum_{p(r),p'(r')} z^{(i)}_p z^{(k)}_{p'} f_{pp'} G_1^2=0$ (the latter is
required by the conditions (39)).

%The eq.(9)  implies that $\rho$ registers the variation of 
%parameters $b$ only through function $y_1$, thus undergoing a 
%single parametric evolution in the matrix space: $\rho(H;b) = \rho(H;y_1[b])$. 
%The functions $y_j(b)$, $j=2 \rightarrow N$ remain constant during 
%the dynamics generated by $L$. 
	
	We find, therefore, that, the diffusion of probability density for ensembles 
of Hermitian operators with  correlated matrix elements (any order), is 
governed by a single parameter $y_1$ with evolution described by eq.(18). 
The eigenvalue statistics of the above ensembles can therefore be studied 
by an exact diagonalization of eq.(18).

\section{Single Parametric Evolution of Eigenvalues}

% The exact diagonalization of the diffusion equation for matrix elements 
% results in a diffusion equation for the eigenvalues $E_i$, $i=1 \rightarrow N$. 

	The eigenvalue equation for the matrix $H$ can be given as $H U = {\underline\lambda} U$ 
with ${\underline\lambda}$ as the diagonal matrix with eigenvalues $\lambda_i$ of $H$ as 
its matrix elements and $U$ as the eigenvector matrix (unitary for complex Hermitian case and 
orthogonal for real-symmetric case). The probability density $P({\underline E},y,b)$  of finding 
eigenvalues $\lambda_i $  between $E_i$ and $E_i+{\rm d}E_i$ at a given $Y$ can then be obtained 
from the matrix elements distribution,
\begin{eqnarray}
P({\underline E},y[b])= \int
\prod_{i=1}^{N}\delta(E_i-\lambda_i) {\tilde\rho (H,y[b])}{\rm d}H 
\end{eqnarray}
Here ${\underline E}$ refers to a diagonal matrix with elements $E_1,..,E_N$. 
 As the $Y\equiv y_1$-dependence of $P$ in eq.(41) enters only through $\rho$, 
a derivative of $P$ with respect to $Y$  can be written as follows   

\begin{eqnarray}
{\partial P\over\partial Y}  
& = &  \int \prod_{i=1}^N \delta(E_i-\lambda_i) 
{\partial {\tilde\rho} \over \partial Y} {\rm d}H 
\end{eqnarray}
The diffusion equation for ${\tilde\rho}$, namely eq.(18), can now be used to 
rewrite eq.(42) as  
\begin{eqnarray}
{\partial P\over\partial Y}  &=& I_1 +I_2 
\end{eqnarray}
where 
\begin{eqnarray}
I_1 &=& \gamma \sum_{\mu} \int   \delta(E-\lambda) 
 {\partial ({H_{\mu} \tilde\rho})\over \partial H_{\mu}} \; {\rm d}H \\
I_2 &=& \sum_{\mu} \int \prod_{i=1}^N \delta(E_i-\lambda_i) 
 {\partial^2 {\tilde \rho}\over \partial H_{\mu}^2}  \; {\rm d}H 
\end{eqnarray}
   
The integral $I_1$ can be simplified by using integration by parts 
 
\begin{eqnarray}
I_1 &=& -\gamma\sum_{\mu} \int \left[{\partial \over \partial H_{\mu}} 
\prod_{i=1}^N \delta(E_i-\lambda_i) \right]\; H_{\mu}\; {\tilde\rho} \;{\rm d}H \\ 
&=& \gamma \sum_{n=1}^N {\partial \over \partial E_n} 
\int \prod_{i=1}^N \delta(E_i-\lambda_i) 
\left[\sum_{\mu} {\partial \lambda_n \over \partial H_{\mu}} 
 H_{\mu}\right] \; {\tilde\rho} \;{\rm d}H  
\end{eqnarray}
A further simplification of the above equation requires a knowledge of 
the rate of change of eigenvalues of $H$ due to a small change in its matrix 
elements. The rate can be obtained by using the eigenvalue equation 
for matrix $H$ along with unitary (or orthogonal for real-symmetric $H$)
nature of its eigenvectors (see appendix A). Using eq.(A2) of the appendix A in 
eq.(47) we get 
\begin{eqnarray}
I_1 &=& \gamma \sum_{n}{\partial \over \partial E_n}\left(E_n P\right)
\end{eqnarray}
The second term can similarly be rewritten as follows
\begin{eqnarray}
I_2 &=&  \sum_n {\partial \over \partial E_n}\sum_{\mu}  {g_{\mu}\over 2}
\int {\partial\over\partial H_{\mu}}\left(\prod_{i}\delta(\mu_i-\lambda_i)
{\partial \lambda_n \over \partial H_{\mu}} \right) {\tilde\rho} {\rm d}H  \\
&=& \sum_{m,n} {\partial \over \partial E_n E_m}
\int\prod_{i}\delta(E_i-\lambda_i)\left[ \sum_{\mu} {g_{\mu}\over 2}
{\partial \lambda_m \over \partial H_{\mu}}
{\partial \lambda_n \over \partial H_{\mu}} \right]\; {\tilde\rho} {\rm d}H
- \sum_{m} {\partial \over \partial E_n}
\int\prod_{i}\delta(E_i-\lambda_i)\left[ \sum_{\mu} {g_{\mu}\over 2}
 {\partial^2 \lambda_n \over \partial H_{\mu}^2}\right]
\; {\tilde\rho} {\rm d}H
\end{eqnarray}
Now by using eqs.(A3,A5) of the appendix A, $I_2$ can be expressed  
in terms of eigenvalue derivatives of $\rho$,  
\begin{eqnarray}
I_2&=& \sum_n {\partial^2 P \over \partial E_n^2}
+ \sum_n {\partial \over \partial E_n}
\left[ \sum_{m \not= n}{\beta P\over {E_m -E_n}}\right]
\end{eqnarray}
A substitution of $I_1$ and $I_2$, given by eqs.(47,51), in eq.(43) leads 
an equation describing the single parametric evolution of the eigenvalues of 
ensemble $\rho(H)$, 

\begin{eqnarray}
{\partial P \over\partial Y} &=& 
\sum_{n}{\partial \over \partial E_n}\left[{\partial \over \partial E_n}
+\sum_{m\not=n}{\beta \over {E_m-E_n}}
+ \gamma E_n \right] P
\end{eqnarray}
%
% (with $Y\equiv y_1$, the notation changed for an easy reference).
The eq.(52) describes the evolution of the eigenvalue density 
$P({\underline E},y[b])\equiv P({\underline E},Y,y_2,..y_M)$ due to variation 
of the parameter $Y$ from an arbitrary initial state, say 
$P({\underline E_0},y[b_0])\equiv P({\underline E_0},Y_{0},y_2,..y_M)$ occurring 
at $Y=Y_{0}$. Note here the parameters $y_j$ ($j>1$) being constants of motion, 
have a same value for both initial ensemble $\rho(H_0,b_0)$ as well as $\rho(H,b)$. 
The evolution of the eigenvalues tends to an steady state 
in limit ${\partial P\over\partial Y} \rightarrow 0$ or $Y\rightarrow \infty$.
The solution of eq.(52) in the limit is a Wigner-Dyson ensemble \cite{alh}: 
$P({\underline E})=\prod_{i<j} |E_i-E_j|^{\beta} {\rm e}^{-{\gamma\over 2}\sum_k E_k^2}$ 
(thus a GOE for $\beta=1$ and a GUE for $\beta=2$). Note, under certain conditions, the steady state 
solution may also correspond to the eigenvalue distribution of an ensemble of anti-symmetric 
Hermitian matrices. A knowledge of the solution of eq.(52) can now help us in determining the 
$n$-level density correlations $R_n$, defined as 
$R_n(E_1,E_2,..E_n;Y)= {N!\over (N-n)!}\int \prod_{j=n+1}^N{\rm d}E_j P({\underline E};Y-Y_0)$. 
The first order correlation $R_1$ is also known as mean level density and its inverse 
gives the mean level spacing $\Delta$ of the full spectrum (that is, averaged of spacings in 
full length of the spectrum).  
By a direct integration, eq.(52) can also be used to study the evolution of $R_n$  
with changing complexity of the system.

	Equation (52) is applicable for arbitrary values of the coefficients 
$b$; it is therefore valid  for the case of uncorrelated Gaussian ensembles too. 
 The latter have been shown to be good models for non-interacting systems \cite{alh}. 
Within random matrix framework, 
therefore, we find that the energy levels of both systems, interacting as well as 
non-interacting,  undergo a same diffusion process with changing system parameters. 
As a result, the level-statistics (and related physical properties) in the two cases 
can be described by the same mathematical formulation. However note that, 
due to different complexity parameters in general, the rate 
of evolution is different in the two cases.

The advantages of a single parametric formulation of the evolution of eigenvalues is 
manifold and have been discussed in detail in \cite{ps1} in context of non-interacting systems
(or uncorrelated ensembles); the similarity of eq.(52) with that of eq.(17) of \cite{ps1} 
allows us to use the discussion given in sections I.(D),I.(E), II of \cite{ps1} for the 
correlated ensembles too (with $\mu$ replaced by $E$). 
However, for completeness sake, we briefly  review it here again.  
  The eq.(52) is similar to the equation governing the evolution 
of eigenvalues in the Dyson's Brownian motion model which was originally introduced by Dyson to 
describe the eigenvalue-dynamics of an ensemble of Hermitian matrices subjected to random 
perturbation; the non-equilibrium states of this model are known as Brownian 
ensembles (see chapter 8 of \cite{meta}). Later on it was shown that 
when an ensemble $H_0$ (fixed or random) is subjected to a random perturbation, of 
strength $\sqrt{Y-Y_0}$, by a standard random matrix ensemble $V$ (described by a 
probability density ${\rm e}^{-(\gamma/2) {\rm Tr} V^2}$), the resulting ensemble 
$H=H_0+(\sqrt{Y-Y_0}) V$ behaves like a Brownian ensemble (BE) (see chapter 14 of \cite{meta}, 
chapter 6 of \cite{fh} and \cite{ap,ps1}).  Here $H_0$ and $V$ may belong 
to a same symmetry class, with $\sqrt{Y-Y_0}$ governing the parametric eigenvalue dynamics, or  
different symmetry classes with $\sqrt{Y-Y_0}$ as a parameter for symmetry admixing transitions. 
The statistical properties of BE  depend only on the parameter $\sqrt{Y-Y_0}$ besides 
underlying symmetry and  many of their correlations are already known\cite{ap}. 
 
  As discussed in \cite{ap}, the mean level density $R_1$ of a BE changes from  an initial state to
a semi-circular form (typical of Wigner-Dyson ensembles) at the scale of 
$\gamma (Y-Y_0) \simeq N \Delta^2_l$ with $\Delta_l(E,Y)$ as the local mean level spacing at energy $E$; 
its evolution can therefore be described in terms of the parameter $(Y-Y_0)$. 
However the transition of level-density correlations to equilibrium, with $(Y-Y_0)$ as 
the evolution parameter, is rapid, discontinuous for infinite dimensions of
matrices \cite{ap}. For small-$Y$ and large $N$, a smooth crossover
can be seen in terms of a rescaled parameter $\Lambda(E)$:
\begin{eqnarray}
\Lambda(E)= \gamma |Y-Y_0|/\Delta_{l}^2
\end{eqnarray}
The limits $\Lambda\rightarrow 0$ and $\Lambda\rightarrow \infty$ correspond to 
the level-statistics approaching the initial state and Wigner-Dyson ensembles, 
respectively. As obvious from the definition of $\Lambda$, an intermediate state between two 
limits occurs when the perturbation $\sqrt{Y-Y_0}$ 
mixes levels in a finite energy-range of $r$ local mean level-spacings, 
$r \simeq \sqrt {Y-Y_0}/\Delta_{\eta}$, $0<r<N$.    
For finite size BE, $\Lambda$ varies smoothly with changing $Y-Y_0$
which results in a continuous family of BEs, parameterized by $\Lambda$. 
However the level-statistics for the large
BE (size $N \rightarrow \infty$) can be divided into three regions:

(i) {\bf initial regime: {$(Y-Y_0) \Delta^{-2}_{l} \rightarrow f(N^{-1})$}}:
If the local mean level spacing $\Delta_{l}$ increases with size $N$ at a rate 
faster than that of $\sqrt{ Y-Y_0}$, the perturbation will mix fewer number of 
levels as system size increases. The level-statistics therefore  approaches its 
initial state in the infinite size limit. 

(ii) {\bf WD regime: $(Y-Y_0) \Delta^{-2}_{l} \rightarrow f(N)$}:
 Due to change in $\Delta_{\eta}$ with size $N$ being slower than that of $\sqrt {Y-Y_0}$), 
even a small change in complexity parameter in this case is capable of mixing the levels 
 in an increasing energy range of many local mean level-spacings. This results in an 
increasing delocalization of eigenfunctions and Wigner-Dyson behavior of level-statistics.
                                                                                    
(iii){\bf Critical regime: $(Y-Y_0) \Delta^{-2}_{l}=f(N^0)=\alpha$= a constant}:
The perturbation in this case mixes only a finite (non-zero), fixed number of levels 
even when the 
system is growing in size. The finite, non-zero $\Lambda$-value in limit
$N\rightarrow \infty$ therefore gives rise to a third statistics,
intermediate between initial ensemble and Wigner-Dyson ensemble, which is known as
the critical Brownian ensemble (CBE). This being the case for arbitrary
values of $\alpha$ (non-zero and finite), an infinite family of critical BE,
characterized by can  occur during initial ensemble $\rightarrow$ Wigner-Dyson 
ensembles.

%Following the equivalence of their 
%diffusion equations, the eigenvalue correlations of the correlated Gaussian 
%ensembles can be obtained just by replacing $\lambda$ by $Y$ in the 
%BE correlations. 

The same evolution equations of $P$ for correlated ensembles and BE imply a similarity
in their eigenvalue distributions for all $Y$-values,
under similar  initial conditions (that is, $P({\underline E},Y_0)$ same for
both the cases). As a result, one obtains the analogous evolution equations for
their correlations $R_n$ too (see eq.(16) of \cite{ap}).  The mean level density $R_1$ of
a correlated ensemble can therefore be given by the mean level-density of a BE with same
$(Y-Y_0)$ value and belonging to a same symmetry class (as that of correlated ensemble).
Further the analogy of evolutions of higher order correlations ($n>1$) in the two cases implies
(i) the discontinuity of transition of $R_n$  for infinite size correlated ensembles,
(ii) a smooth crossover of $R_n$ for finite correlated ensembles.
The crossover parameter for correlated ensembles can again be defined by eq.(54) 
where now $Y-Y_0$ is the complexity parameter of the correlated ensemble and 
$\Delta_{l}$ as its local mean level spacing. 
Note, in the case of $d$-dimensional disordered systems of linear size $L$, the number of 
states in 
a volume of linear dimension $\zeta$ in $d$-dimensions is $n(0)\zeta^d$ with $n(0)$ 
as the density of states at Fermi level and $\zeta$ as the localization length. 
Consequently, the typical energy separation between 
such states  is $\Delta_{l}(E,Y)=(n(0)\zeta^d)^{-1}$. Similarly 
the mean level spacing of states in the full length of the spectrum 
is $\Delta(E,Y)=(n(0) L^d)^{-1}$. For disordered systems, the local mean level 
spacing $\Delta_l$ can therefore be expressed in terms of the mean level density $R_1$ 
as $\Delta_l=(L/\zeta)^d R_1^{-1}$.

%with  $\eta\equiv \eta(E,Y)$ as the correlated "volume" and the 
%$\Delta\equiv \Delta(E,Y)=R_1^{-1}$ 
%as the mean level spacing of the whole spectrum at energy $E$ and the
%parameter $Y-Y_0$. 
%
%The correlation volume $\eta$ is a measure of the degree of
%localization of the eigenfunctions and can be determined by a knowledge
%of the inverse participation ratio $I_2$; $\eta \propto (I_2)^{-1}$.
%

%Note, as in the BE case, a prior rescaling of the 
%parameter $Y$ is required to see a smooth crossover for high energies or large but finite 
%system sizes (see \cite{ap,ps1,ps2} or chapter 14 of \cite{meta}  
%for details in the BE case). The rescaled parameter is given 
%by $\Lambda(E,Y)= (Y-Y_{0})/D_{\eta}^2$, with $D_{\eta}(E)$ as the local mean level spacing 
%at energy $E$, $D_{\eta}=D N/\eta$, $D$ as the mean level spacing of the full spectrum and 
%$\eta$ as the correlation volume around energy $E$\cite{ps1}. 
%The limits $\Lambda\rightarrow 0$ and $\Lambda\rightarrow \infty$ correspond to 
%the level-statistics approaching the initial state and Wigner-Dyson ensembles, 
%respectively. As obvious from the definition of $\Lambda$, an intermediate state between two 
%limits occurs when the perturbation $\sqrt{Y-Y_0}$ 
%mixes levels in a range of $n$ mean level-spacing, $n \simeq {\sqrt Y-Y_0}/D_{\eta}$.    

	The parameter $\Lambda$, being a function of the distribution parameters of the 
matrix elements, is sensitive to the changes in the system parameters, due to their influence 
on the system-interactions and their uncertainties. Some examples of such system parameters 
are disorder, dimensionality, boundary and topological conditions, system size etc. 
The presence of disorder randomizes the interactions in the system with degree of disorder 
 affecting the distribution parameters $b$ and consequently $\Lambda$. The dependence 
of $\Lambda$ on dimensionality and boundary conditions can be explained by using a simple 
example. Consider a $N\times N$ lattice with a Gaussian site disorder as well as 
a Gaussian type random interaction between nearest-neighbor sites. The lattice 
Hamiltonian $H$, in site representation, is a sparse matrix with only $(Z+1)N$ non-zero, 
independently distributed matrix elements; here $Z$ is the number of nearest-neighbors of 
a site. Consequently only $(Z+1)N$ $b$-parameters (out of $N^2$) contribute to  $Y$. As the 
coordination number $Z$ is different for different dimensions and boundaries of the lattice, 
the $Z$-dependence of $Y$  results in its dependence on the dimensionality as well as 
boundary conditions of the system. 
Further the local mean level spacing is also sensitive  to the dimensionality as well as the 
boundary conditions.  A variation of any of the latter parameters can affect both $\Delta_l$ and $Y$ 
and therefore   $\Lambda$; (See also \cite{ps2} where the dependence of $\Lambda$ on 
system parameters is explained  by considering an example of Anderson Hamiltonian). 

	The size-dependence of $\Lambda$ also plays a crucial role in determining the 
level statistics of the correlated ensemble in the crossover regime. In general, 
both  $Y-Y_0$ as well as the local mean level density are the functions of  
system-size $N$ which results in $N$-dependence of $\Lambda$. 
As a consequence, the level statistics in finite systems smoothly approaches one of the 
two end points, namely, $\Lambda\rightarrow 0$ or 
$\Lambda \rightarrow \infty$, with increasing system size. 
However, as in BE case, the variation of $\Lambda$ in infinite correlated ensembles  
may lead to an abrupt transition, 
with its critical point given by the condition $\Lambda=\it {size \; independent}$. 
As in the BE case,  the size-independence of $\Lambda$ at the critical point, 
results in a level-statistics different from the two end points. 
Note if the size-dependence of $\Delta_{l}^2$ in a correlated ensemble remains different 
from that of ${Y-Y_0}$ under all complexity conditions, the system will never undergo 
a transition in level-statistics. 

\section{Examples}

	In this section, we consider two examples corresponding to $2^{nd}$ and $3^{rd}$ 
order matrix elements correlations and provide the theoretical formulation for $2$-point 
eigenvalue correlations for the cases by using the Brownian ensemble analogy.  

\subsection {Quantum Hall System} 

 Let $H=H_0+V({\bf r})$ be the 
single particle Hamiltonian for a disordered quantum Hall system 
with $H_0$ as the kinetic energy of the electrons and $V({\bf r})$  as a space correlated 
disordered potential e.g 
$<V({\bf r}) V({\bf r'})>=f({\bf r},{\bf r'})$ (with $<V({\bf r})>=0$). 
Using the Landau states $\psi_{nk}(r)\equiv <{\bf r}|nk>$  
(the eigenstates of $H_0$) as the basis, $H$ can be written as 
$H_{nk;n'k'} =\epsilon_n \delta_{n,n'}\delta_{k,k'} +V_{nk;n'k'}$
where $H_{nk;n'k'} \equiv <nk|H|n'k'>$ and $\epsilon_n=(n+1/2)\hbar \omega$ as 
the eigenvalues of $H_0$. The interaction between impurities  
results in a correlation of the matrix elements of $V$ and thereby $H$ \cite{bh}:
\begin{eqnarray}
<V_{n_1 k_1 ;n_2 k_2} V_{n_3 k_3; n_4 k_4}>
=\int {\rm d} {\bf r} {\rm d} {\bf r'} \psi_{n_1 k_1}^*({\bf r}) 
\psi_{n_2 k_2}({\bf r}) 
\psi_{n_3 k_3}^*({\bf r'}) \psi_{n_4 k_4}({\bf r'}) f({\bf r},{\bf r'}).
\end{eqnarray}

The parameter $\Lambda$ can now be determined if mean level spacing $\Delta$ and the  
real-space correlations for the potential $V$ are explicitly known. For example, 
consider the case when magnetic field $B$ becomes much stronger than 
the disorder  potential. The Hamiltonian matrix $H$ in this case is divided into 
various independent blocks (each corresponding to a different Landau level) and the 
statistics of energy states in each  Landau Level can be discussed independently 
\cite{bh}. For a Gaussian type disorder
%\begin{eqnarray}
$ <V({\bf r}) V({\bf r'})>=
(V_0^2 /2\pi \sigma^2){\rm e}^{-|{\bf r}-{\bf r'}|^2/2\sigma^2}$,
%\end{eqnarray}
the matrix element correlations in the lowest Landau level $n=0$ can be given as
$< V_{0 i;0 j}\; V_{0 k;0 l} > \equiv < V_{ij}\; V_{kl} >$ where 
\begin{eqnarray}
< V_{ij} V_{kl} > = (V_0^2/l_c L_y \alpha \sqrt{2\pi})
\delta(i-j,l-k) {\rm e}^{-(i-j)^2 \alpha^2 /2}
{\rm e}^{-(i-k)^2 /2\alpha^2}
\end{eqnarray}
with $\alpha^2=(1+\sigma^2/l_c^2)$ as a measure of the correlation length of the potential 
relative to the magnetic length $l_c=({\hbar/eB})^{1/2}$\cite{bh}. 
Using the notation $H_{0 k;0 l}\equiv H_{kl}$, the  distribution parameters of the 
matrix elements of the Hamiltonian in the Landau level $n=0$ can be given as 
\begin{eqnarray}
 <H_{ij}>  &=& \epsilon_0 \delta_{ij} \\
< H_{ij;s}\; H_{kl;s} > &=& \epsilon_0^2 \delta_{ij}\delta_{kl} 
+ < V_{ij;s} V_{kl;s} >
\end{eqnarray}
with $<V_{ij;s} V_{kl;s} >$ as the correlations between different components of the 
elements of $V$,
\begin{eqnarray}
< V_{ij;s}\; V_{kl;s'} > = {1\over 2} [<V_{ij} V_{lk}>+(-1)^{s-1} <V_{ij} V_{kl}>]\; \delta_{ss'}
\end{eqnarray}
The distribution of the local Hamiltonian for the lowest Landau level can then be 
represented by eq.(1) with parameters $b$ obtained from eqs.(56,57) 
(here $\mu_1 \equiv (ij;s)$, $\mu_2 \equiv (kl;s)$); see appendix C for an example.  
A substitution of the $b$-parameters in eq.(15) gives us the complexity parameter 
governing the energy level dynamics in the lowest Landau level. As shown in the appendix C 
by  a simple case $N=2$, the parameters $b$ and, therefore $Y$, turns out to be 
 a function of the disordered potential $\alpha$, $V_0$, system length $L_y$ as well as 
magnetic field $B$ and can be varied by changing any one of them. 

The above discussion is valid for higher order landau Levels too, with complexity 
parameter still described by eq.(15) however the coefficients $b_{\mu_1\mu_2}$ are different 
for different Landau levels (see \cite{bh} for the matrix element correlations of potential $V$). 
 The rate of transition of level-statistics  therefore differs, in general, from one Landau level 
to another. For weak magnetic fields, where various Landau levels can not be considered 
as independent, $H$ can still be represented by the ensemble eq.(1). However, now the number of 
coefficients $b_{\mu_1\mu_2}$ which contribute to $Y$ is much larger (due to correlations 
between levels in two different Landau Levels).
 
%The dynamics of the energy levels in higher order Landau levels can also be described 
%by the 

% Note that $b_{ijkl}$ is also a function of many constants, that is, basis indices $i,j,k,l$ 
%etc and parameters 
% $Y_2, Y_3,..Y_M$ can be identified with various combinations of these constants. 

%For notational clarification, however, let us rewrite  
%\begin{eqnarray} 
%\rho(H)={\rm exp}\left[{-\sum_{s}\sum_{i,j;i\le j}
% \sum_{k,l;k\le l} b_{ijkl;s}  H_{ij;s} H_{kl;s}}\right]  
%\end{eqnarray}
% with $H_{kl}\equiv <nk|H|nl>$ and 
%$<H_{ij;s} H_{kl;s}>={\partial {\rm log} C \over \partial b_{ijkl;s}}$.
%A single parametric formulation  of the eigenvalue evolution in this case can 
%be obtained by using a transformation $b\rightarrow y$ which satisfies following condition: 
% 
%\begin{eqnarray}
%\sum_{ijkl;s} f_{ijkl;s}{\partial  y_m\over \partial b_{ijkl;s}}= \delta_{m1}
%\end{eqnarray}
%where $f_{ijkl;s}= \gamma b_{ijkl;s}-\sum_{mn} b_{ijmn;s} b_{mnkl;s}$.  
% The complexity parameter $Y$ can now be given by 
% $Y= \sum_{i,j,k,l;s} \int {{\rm d} z_{ijkl} b_{ijkl;s} G^{-1} }
% +constant $
%where $G=\sum_{ijkl;s} z_{ijkl} f_{ijkl;s} $ 
% with $z_{ijkl}$ as functions which satisfy the condition 
%$\sum_{ijkl;s} z_{ijkl}\sum_{mn}b_{ijmn;s}b_{mnkl;s} = 0$.    

% and $\tilde N$ as the number of nonzero elements in set $b\equiv \{b_{ijkl;s}\}$. 

In absence of disorder, under the independent Landau Level approximation,  all energy levels 
in a given Landau Level are degenerate and matrix $H=H_0$ is a diagonal matrix with a Poisson
behavior for its eigenvalues (due to dominance of zero spacings).  The switching 
of disorder removes the degeneracy and delocalizes the wavefunction if the impurities 
are interacting. The degree of delocalization depends on the strength of impurity interactions 
with respect to the magnetic field strength $B$. 
If the latter is strong enough to mix the levels in an energy range of many mean level spacings,  
(which corresponds to the limit $H\approx V$),  the energy levels of $H$ show a GUE behavior.
(This is similar to the case of strongly interacting many body systems e.g. the statistics of 
resonances in complex nuclei which can be well-modeled by GOE or GUE \cite{alh}).  
%The variation of parameter $\alpha$ subjects, therefore, the system to undergo a cross-over 
%from Poisson ($\alpha=0$) to GUE behavior ($\alpha=
Under an intermediate state of disorder, therefore, the ensemble H lies 
between the Poisson ensemble and GUE 
and can be modeled by eq.(1). The level statistics for this case can then be given 
by the one for a BE appearing during a $Poisson \rightarrow$ GUE transition . 
The two point correlator $R_2$ \cite{ap} for states in lowest Landau level can therefore 
be given as
\begin{eqnarray}
 R_2 (r;\Lambda) - R_2(r;\infty)={4\over \pi}\int_0^\infty {\rm d}x 
\int_{-1}^1 {\rm d}z \;{\rm cos}(2\pi rx) 
\;{\rm exp}\left[-8\pi^2\Lambda x(1+x+2z\sqrt x)\right]
\left({\sqrt{(1-z^2)}(1+2z \sqrt x) \over 1+x+2z \sqrt x}\right)
\end{eqnarray}
where $R_2(r,\infty)=1-{{\rm sin}^2(\pi r)\over \pi^2 r^2}$ (the GUE limit); the above 
formulation was obtained for the BE in \cite{ap}.

Let us now consider the case with weak magnetic fields where the interaction between various 
Landau levels can not be ignored. The eigenvalue spectrum of $H_0$ for this case behaves 
as a uniform spectrum \cite{bh} (that is, an initial spectrum of uniform spacing). 
The switching of interacting-impurities potential $V$ 
again results in broadening of the wavefunctions. In the limit where impurity interactions 
are strong enough to 
mix energy levels in different Landau levels (that is $H \approx V$), the eigenvalues of 
$H$ show a GUE behavior. The varying degree of the interaction between impurities therefore 
leads to a transition of the level statistics from uniform spectrum $\rightarrow$ GUE behavior. 
The two point correlation for the level-statistics at any intermediate stage of 
impurity interaction can then be given by that of a BE appearing during 
the uniform spectrum $\rightarrow$ GUE transition:

\begin{eqnarray}
 R_2 (r;\Lambda) - R_2(r;\infty)= 2\sum_{q=-\infty}^{\infty} {\rm e}^{-8\pi^2 q^2\Lambda}
\int_{0}^1 {\rm d}x (1-x) {\rm e}^{-8\pi^2 qx \Lambda} \;
\left[{\rm cos}(2\pi qr)-{\rm cos}(2\pi (q+x)r)\right] 
\end{eqnarray}
with $\Lambda \rightarrow \infty$ corresponding to GUE limit.

\subsection{\bf Disordered Systems with fermionic interactions} 

Consider a general Hamiltonian for spinless interacting fermions

\begin{eqnarray}
H=\sum_{ij} V_{ij} a_i^+ a_j + 
{1\over 4} \sum_{ijkl} U_{ijkl} a_i^+ a_j^+ a_l a_k
\end{eqnarray}

here the states $|i>=a_i^+|0>$ describe a fixed basis 
of $m$ single-particle states with $V_{ij}$ as  matrix elements of 
the one body Hamiltonian and $U_{ijkl}$ as the antisymmetrized matrix 
elements of the two body interaction $U$. 
The presence of disorder randomizes both $V$ and $U$.
For example, for $V$ as a white noise, its matrix elements are independently 
distributed random variables. However the fermionic interaction results in correlations 
among matrix elements of $U$:
\begin{eqnarray}
< U_{ijkl} U_{klmn} U_{mnij} >
=\int {\rm d}^3 r_1 {\rm d}^3 r_2  {\rm d}^3 r_3
\psi_{ij}^*(r_1) \psi_{kl}(r_1)
\psi_{kl}^*(r_2) \psi_{mn}(r_2)
\psi_{mn}^*(r_3) \psi_{ij}(r_1)
f(r_1,r_2) f(r_2,r_3) f(r_3,r_1)
\end{eqnarray}
with $f({\bf r_1},{\bf r_2})$ as the interaction between two fermions at positions 
${\bf r_1}$ and ${\bf r_2}$.
The Hamiltonian $H$ will therefore be a matrix with varying degree of correlations 
between its elements and can be represented by ensemble eq.(19). 
In absence of fermionic interaction, the matrix $H=V$ and its 
statistical properties depend on the degree of disorder. For example, in presence of strong 
disorder, the eigenvalues  of $V$ show a Poisson distribution with localized eigenfunctions 
\cite{ps2,gw,alh}.  
The weak disorder limit of $V$ shows a Wigner-Dyson distribution for its eigenvalues with 
extended eigenfunctions. Similarly in absence of disorder, $H=U$ and its statistical behavior 
is governed by the fermionic density in various parts of the system. 
The presence of almost uniform fermionic density in the system leads to delocalization 
of eigenfunctions and  $U$ behaves like an ensemble of anti-symmetric Hermitian matrices 
(see \cite{meta} for details on anti-symmetric random matrices). 
However non-uniform electronic density in various parts of the system   
(that is, stronger interactions in certain parts of the system as 
compared to others) can result in localization of wavefunctions and 
thereby a Poisson behavior for the eigenvalues of $U$.   

In presence of both disorder as well as fermionic interaction, 
the behavior of $H$ is governed by the inter-competition 
between them. In this case, it is  preferable to represent $H$ in the 
$N=m^2$ dimensional basis of two particle states $|ij>$; the choice of the basis 
results in appearance of the disorder elements as the diagonal elements of $H$ and 
fermionic interaction elements as the off-diagonals: $<ij|H|ij>=V_{ij}$ and 
$<ij|H|kl>=U_{ijkl}$. The changing strength  of the fermionic interaction 
in presence of disordered potential subjects the level-statistics to undergo a 
transition from the initial state (given by the statistics of $V$)  to Wigner-Dyson 
statistics (when $U \simeq V$). However, in the limit when the disorder potential 
becomes negligible as compared to fermionic interactions, the level-statistics of 
$H$ approaches that of anti-symmetric matrices. As obvious, the level-statistics 
for all other cases, corresponding to different strengths of disorder potential and 
fermionic density, 
will lie on the transition curve from Poisson $\rightarrow$ Wigner-Dyson $\rightarrow$ 
anti-symmetric ensembles. Using eq.(19) as the model for intermediate states, the 
level-statistics for this case can then be described by BE lying between 
Poison $\rightarrow$ anti-symmetric ensemble.

        The ground state properties of electrons in nano-particles or quantum 
dots, that is, finite systems of fermions interacting via coulomb forces, are not  
 yet fully understood. Using ensemble (19) or (32) as their model, the physical 
properties of these system can now be probed further and earlier experimental observations 
may be explained. For example, it has been experimental observed that the  peak-spacing 
statistics for an irregular quantum dot in Coulomb blockade regime undergoes a crossover 
from Wigner-Dyson to Gaussian behavior as the strength of  electron-electron 
interaction increases\cite{us,srp}. Within our formulation, the observed behavior can 
be explained as follows. The single electron dynamics inside a quantum dot of irregular 
shape is chaotic; the level-statistics of single particle Hamiltonian $V$ can therefore be 
modeled by the Wigner-Dyson behaviour \cite{gw}. The addition of more electrons switches 
the potential $U$, however, due to non-uniform electronic density during initial stages, the 
correlation between various matrix elements of $U$ need not be same. 
The statistical behavior of the quantum dot 
can therefore be described by an ensemble given by eq.(19). As electron density increases, $U$ 
dominates over $V$ and the level-statistics approaches the behavior of anti-symmetric 
matrices. It is already known that the nearest-neighbor spacing distribution for 
non-central spacings in the spectrum of anti-symmetric Hermitian matrices behaves 
like a Gaussian distribution \cite{meta}. The observed Gaussian behavior of the peak-spacings 
in the strong interaction limit is therefore well in agreement with theoretical 
expectations. Using ensemble (19) as a model for the quantum dot Hamiltonian, the behavior of 
peak-spacings in the intermediate regime can be predicted to be similar to that of a BE appearing 
during a cross-over from GOE $\rightarrow$ anti-symmetric ensembles. 
A detailed quantitative analysis of such cases is still in progress and will be published 
elsewhere.

	In general, the size dependence of the parameter $Y-Y_0$ and the local mean level 
spacing $\Delta_l$ for a system with e-e interaction is different from the non-interacting 
systems. In interacting systems, therefore, the critical point of level-statistics, 
given by condition $\Lambda={\it N-independent}$, can occur at a disorder 
strength (or energy) different from the one for non-interacting systems. This is consistent 
with the results given by renormalization group techniques \cite{mm} which show 
that the introduction of interactions into quantum dots can produce 
phase transitions in the limit of weak disorder, leading to behavior qualitatively 
different from the non-interacting case. 
Further for one dimensional non-interacting systems, it is known that the wavefunctions are 
localized even in a weak disorder limit. However our formulation indicates that the 
fermionic interaction may lead to extended states even in one dimensional disordered 
system; the implication is in agreement with earlier studies in this context\cite{du}.    

%....

\section{Conclusion}

   In summary we show, for the first time, that the level-statistics of disordered systems 
with interactions is governed by a single parameter, namely, the rescaled 
complexity parameter $\Lambda$. Note the level-statistics of non-interacting systems can 
also be described in terms of $\Lambda$ \cite{ps1}. However the introduction of interactions 
modifies the dependence of $\Lambda$ on system parameters which can significantly affect the 
location of the critical point for the phase transitions and corresponding 
level statistics.  Our study also reveals a deep level of universality 
underlying physical systems, namely, the  Brownian ensembles as the statistical back bone of  
both interacting as well non-interacting systems. This universality should be explored in 
full detail as it may reveal many new connections among a wide range of complex systems and 
can be helpful in theoretical formulation of many of their physical properties.

%We find, that similar to non-interacting case, the level statistics of 
% disordered systems with interactions can be mapped to single parametric Brownian ensembles. 
%However the introducton of interactions changes the universality class of Brownian 
%ensemble model which affects the location of the critical point, and, thereby corresponding 
%level statistics.   

%The statistics of energy levels, and therefore 
%wavefunctions, in disordered systems seems to depend on the nature and 
%relative strength of interactions with respect to disorder. 
%As the varying degree of matrix element 
%correlations may result in an intermediate statistics between Poisson and 
%SGE (the two being indicators of  fully localized and  
%delocalized wavefunctions, respectively), 
%this implies the following: 
%the presence of local interactions can lead to a localization of 
%wavefunctions even in a weak disorder limit. On the other hand, 
%the interactions homogeneous in nature (of same order throughout the 
%system) and stronger than disorder can delocalize the wavefunctions. 

\begin{appendix}

\section{The Change of Eigenvalues and Eigenfunctions} 

	The eigenvalue equation of a complex Hermitian matrix $H$ is given by 
$H U = U \Lambda$ with $\Lambda$ as the matrix of eigenvalues $\lambda_n$ 
and $U$ as the eigenvector matrix, unitary in nature. As obvious, a slight variation of the 
matrix elements of $H$ will, in general, lead to variation of both the eigenvalues as well as 
the eigenvectors and  associated rates of change can be obtained as follows 
(see appendices A-E of \cite{ps1} for more details): 

As $\lambda_n = \sum_{i,j} U_{ni} H_{ij} U_{nj}^*$,  
 the rate of change of $\lambda_n$ 
with respect to $H_{kl;s}$ (with $s$ referring to real, $s=1$, and imaginary, 
$s=2$, parts of $H_{kl}$) can be given  

\begin{eqnarray}
{\partial \lambda_n \over \partial H_{kl;s}}= 
   {i^{s-1}\over g_{kl}} [U_{ln} U_{kn}^{*} - (-1)^s U_{ln}^{*} U_{kn} ].  
\end{eqnarray}

where $g_{kl}=1+\delta_{kl}$. This
 can further be used to obtain the following relations

\begin{eqnarray}
\sum_{k\le l}
 \sum_{s=1}^2 {\partial \lambda_n \over\partial H_{kl;s}} H_{kl;s}  
&=& \sum_{k, l}  
 H_{kl} U_{ln} U_{kn}^{*} = \lambda_n 
\end{eqnarray}
and 
\begin{eqnarray}
\sum_{k\le l} g_{kl}
 \sum_{s=1}^2 {\partial \lambda_n \over\partial H_{kl;s}}   
  {\partial \lambda_m \over\partial H_{kl;s}}  
 = 2\delta_{mn}
\end{eqnarray}

As obvious from eq.(A1),  the second order  
change of an eigenvalue with respect to a matrix element requires a knowledge 
of the rate of change of one of the eigenvector components with respect 
to $H_{kl}$. The latter can again be obtained by using the eigenvalue equation,   

\begin{eqnarray}
{\partial U_{pn} \over\partial H_{kl;s}}=
{i^{s-1} \over g_{kl}} \sum_{m\not=n}
{1\over {\lambda_n -\lambda_m}}
U_{pm}(U^{*}_{km}U_{ln} + (-1)^{s+1} U^{*}_{lm} U_{kn})
\end{eqnarray}

Now by differentiating eq.(A1) with respect to $H_{kl;s}$ and by using the eq.(A4) 
we can show that

\begin{eqnarray}
\sum_{k\le l} g_{kl} 
 \sum_{s=1}^2 {\partial^2 \lambda_n \over\partial H_{kl;s}^2}   
&=& 2\beta \sum_{m} {1 \over \lambda_n - \lambda_m}
\end{eqnarray}

	For the real-symmetric case, the corresponding relations 
can be obtained by using $U^+=U^T$ (as eigenvector matrix is 
now orthogonal) in eq.(A1) and taking $H_{ij;2}=0$ for all 
values of ${i,j}$. 

\section{Solution of Eq.(38)}
 
According to theory of partial differential equations (PDE)
\cite{sne},
the general solution of a linear PDE
\begin{eqnarray}
\sum_{i=1}^M P_i(x_1,x_2,..,x_M){\partial Z \over \partial x_i} = R  
\end{eqnarray}
is $F(u_1,u_2,..,u_n)=0$ where $F$ is an arbitrary function and
$u_i(x_1,x_2,..,x_n,Z)=c_i$ (a constant), $i=1,2,..,n$ are independent
solutions of the following equation
                                                                                
\begin{eqnarray}
{{\rm d}x_1 \over P_1} =
{{\rm d}x_2 \over P_2} =.....
{{\rm d}x_k \over P_k} =......
{{\rm d}x_M \over P_M} =
{{\rm d}Z \over R}
\end{eqnarray}

Note the function $F$ being arbitrary, it can also be chosen as 
\begin{eqnarray}
F\equiv\sum_j (u_j-c_j)=0  
\end{eqnarray}

	The equations for various $y_j$ in  the set of eq.(38) are of the 
same form as eq.(B2) and, therefore, can be solved as described above. 
Let us first consider the equation for $y_1$; its general solution can be given 
by a relation $F(u_{1},u_{2},..,u_{M}) = 0$
where  function $F$ is arbitrary and $u_j$ are the functions of $M$ parameters 
of set $b$ such that $u_{j}(\{b\},y_1)=c_j$ (with $c_j$'s as constants). The functions  
$u_{j}$ are the independent solutions of the equation
                                                                                
\begin{eqnarray}
{{\rm d}b_{p(1)} \over f_{p(1)}} =....=
{{\rm d}b_{p(2)} \over f_{p(2)}} =.....
{{\rm d}b_{p(r)} \over f_{p(r)}} =......
={{\rm d}y_1}
\end{eqnarray}
where the equality between ratios is implied for all possible combinations $p(r)$ 
of $r$ terms, $r=1\rightarrow n$, with  $M$ as the total number of combinations.  
It is easy to see that each of the above ratios is equal to 
${\sum_{r,p(r)} z_p^{(1)} {\rm d}b_p \over \sum_{r,p(r)} z_p^{(1)} f_p}$ where $z_p^{(1)}$ are 
arbitrary functions. The eq.(B4) can therefore be rewritten as 
\begin{eqnarray}
 {\rm d}y_1={\sum_{r,p(r)} z_p^{(1)} {\rm d}b_p \over \sum_{r,p} z_p^{(1)} h_p} 
\end{eqnarray}
A solution, say $u_1$ of eq.(B5), or alternatively eq.(B4), can now be obtained by 
choosing the functions $z_p^{(1)}$ such that the right side of the above equation 
becomes an exact differential:
\begin{eqnarray}
u_1\equiv y_1 - \sum_{r,p(r)} \int {\rm d}b_{p}  z_{p}^{(1)} G_1
 = constant 
\end{eqnarray}
where $G_1=[\sum_{r,p(r)} z_p^{(1)} h_{p}]^{-1}$.  
The general solution for $y_1$ can therefore be given by a combination of all 
possible functions $u$ obtained by using arbitrary set of $z$-functions. 
%It can be shown that only $M$ of the solutions  would be independent and 
It can be shown that each such solution differ from the other 
only by a constant: $u_j=u_i$ + constant; (this is due to equality of the two ratios
obtained by choosing two different sets $z^{(1)}$ of the functions). 
 The $y_1$ can therefore be written as follows,
\begin{eqnarray}
y_1 = \sum_{r,p(r)} \int {\rm d}b_{p}{z_p^{(1)}  G_1} + constant
\end{eqnarray}
which gives eq.(40) for $k=1$. 

%\begin{eqnarray}
%M y_1 -\sum_{j=1}^M \sum_{r,p(r)} \int {\rm d}b_{p(r)}{z_{pj}^{(1)}  G_j}-constant=0  
%\end{eqnarray}
%which by rearranging the terms gives eq.(). 

The set of equations (38) can similarly be solved  for other $y_j$ ($j>2$). 
For example, the solution of eq.(38) for $y_k$ can be given by the 
function $F_k(v_1,..v_M)=0$ where $v_j(\{b\},y_k)=constant$ are the independent 
solutions of following equality   

\begin{eqnarray}
{{\rm d}b_{p(1)} \over f_{p(1)}} =....=
{{\rm d}b_{p(2)} \over f_{p(2)}} =.....
{{\rm d}b_{p(r)} \over f_{p(r)}} =......
{{\rm d}y_k \over 0} .
\end{eqnarray}
A solution, say $v_1$, of eq.(B8) can now be given as 
\begin{eqnarray}
v_1\equiv y_k - \sum_{r,p(r)} \int {\rm d}b_{p} z_{p}^{(k)} = constant 
\end{eqnarray}
where $z_{p}^{(k)}$ are arbitrarily chosen $M$ functions which satisfy the condition 
\begin{eqnarray}
\sum_{r,p(r)} z_p^{(k)} f_p =0
\end{eqnarray}
As obvious, one possible choice for $z^{(k)}$ functions satisfying  the above condition 
is $z_p^{(k)}=0$ for all $p(r)$ which gives $y_k=constant$.   

As each solution of eq.(B8) is different from the other only by a constant, 
the general solution for $y_k$, $k>1$, can now be given as  
\begin{eqnarray}
y_k = \sum_{r,p(r)} \int {\rm d}b_{p} z_p^{(k)}  + constant 
\end{eqnarray}
The eq.(B7) and eq.(B11) together give the set of eqs.(40).

\section{Example for Quantum Hall Case}

	Within independent Landau level approximation, 
the Quantum Hall ensemble can be described by the probability density 
$\rho(H)$  given by eq.(1) with coefficients $b$ given by eqs.(57,58). However 
it can further be simplified by choosing the origin of energy at $\epsilon_0$ 
which makes $b_{\mu}=<H_{\mu}>=0$. 
The matrix element distribution in the Quantum Hall case can now be 
described by a probability density
\begin{eqnarray}
\tilde{\rho} (H,b)=C{\rm exp}[ -\sum_{\mu_1,\mu_2=1}^M  b_{\mu_1\mu_2}
 H_{\mu_1} H_{\mu_2})] 
\end{eqnarray}
with $C$ as the normalization constant 
\begin{eqnarray}
C^{-1} = \int {\rm d} H {\rm e}^{-\sum_{\mu_1\mu_2}  b_{\mu_1 \mu_2}
 H_{\mu_1} H_{\mu_2})} 
=(\prod_{s=1}^2 {\rm Det} {B_s})^{-1/2}
\end{eqnarray}
Here $B^{(s)}$ is the matrix of coefficients $b_{\mu_1\mu_2}$. 
The parameters $b_{\mu_1\mu_2}$ are related to second 
order correlation $<H_{\mu_1} H_{\mu_2}>$:
\begin{eqnarray}
<H_{\mu_1} H_{\mu_2}> &=& \int H_{\mu_1} H_{\mu_2} {\tilde \rho}{\rm d} H \\
&=& {\eta_{\mu_1\mu_2}\over 2} {\partial {\rm ln} C \over \partial b_{\mu_1\mu_2}}
\end{eqnarray}
where $\eta_{\mu_1\mu_2}=2$ and $1$ for pairs $\{\mu_1\}=\{\mu_2\}$ and 
$\{\mu_1\}\not=\{\mu_2\}$ respectively. 
Let $Q^{(s)}$ be the matrix with its elements as the correlations between 
elements of $H$, that is, $(Q^{(s)})_{\mu_1\mu_2}=(4/\eta_{\mu_1\mu_2}) <H_{\mu_1} H_{\mu_2}>$.  
By using eq.(C2) in eq.(C4), we get    

\begin{eqnarray}
Q^{(s)}_{\mu_1\mu_2} 
&=& {\partial {\rm ln} {\rm Det}[ B^{(s)}]\over \partial b_{\mu_1\mu_2} } 
= {{\rm Cof}(B_{\mu_1\mu_2})\over {\rm Det} [B^{(s)}]} = 
=   {(B^{(s)})^{-1}}_{\mu_2\mu_1}
\end{eqnarray} 
where ${\rm Cof}(B^{(s)}_{\mu_1\mu_2})$ implies the cofactor of the element $B_{\mu_1\mu_2}$ 
in the matrix $B_s$. This implies $Q^{(s)}=(B^{(s)})^{-1}$. The matrix $B^{(s)}$ 
for the QH case can 
therefore be obtained by inverting the correlation matrix $Q^{(s)}=\{Q^{(s)}_{\mu_1\mu_2}\}$. 

Let us consider the case $N=2$. Using the $H_{\mu}$ notation, various components 
of matrix elements can now be denoted as 
$H_1=H_{11;1}, H_2=H_{12;1}, H_3=H_{22;1}, H_4=H_{12,2}$.  
 Following eq.(60), the correlation matrix 
$Q^{(1)}$ in this case is a $3\times 3$ matrix 
\begin{eqnarray}
Q^{(1)} ={\pmatrix {a & o & 2ax \cr \nonumber 
\\ o & a/2 x_1 & 0 \cr  \nonumber \\ 2ax & 0 & a }} 
\end{eqnarray} 
where $a=(V_0^2/2 l_c L_y \alpha {\sqrt{2\pi}})$, 
$x={\rm e}^{-1/2\alpha^2}$ and $x_1={\rm e}^{\alpha^2/2}$. 
By using the relation (C5),  $B^{(1)}$ can be given as  
\begin{eqnarray}
B^{(1)} = {1\over a(1-4 x^2)} {\pmatrix {1 & o & -2 x \cr \nonumber 
\\ o & 8 x_1 (1-4x^2) & 0 \cr \nonumber \\ -2 x & 0 & 1}} 
\end{eqnarray} 
Due to Hermitian nature of $H$, only its off-diagonal elements have imaginary parts. 
For $N=2$ case, therefore, $B^{(2)}$ is just a $1\times 1$ matrix, corresponding to 
correlation $<H_4 H_4>$ with $\mu_4 \equiv (12;2)$:
$B^{(2)}=b_{44}= (2 <H_4 H_4>)^{-1}$. 
The parameter set $y$ for this case can now be obtained by solving the condition 
$A_k=\sum^4_{i,j=1} f_{ij} {\partial y_k\over \partial b_{ij}}
=\delta_{k1}$ where $f_{ij}= \gamma b_{ij} - (1/2) \sum_{k=1}^4 g_k c_{ik} c_{jk}$; 
here $g_k=2$ for odd $k$ and $g_k=1$ for even $k$. As discussed 
in appendix (B), a solution of the condition can be given as 
\begin{eqnarray}
{{\rm d}b_{11} \over f_{11}} =
{{\rm d}b_{12} \over f_{12}} =
{{\rm d}b_{13} \over f_{13}} =
{{\rm d}b_{22} \over f_{22}} =
{{\rm d}b_{31} \over f_{31}} =
{{\rm d}b_{23} \over f_{32}} =
{{\rm d}b_{33} \over f_{33}} =
{{\rm d}b_{44} \over f_{44}} =
{{\rm d}y_k \over \delta_{k1}}.
\end{eqnarray}
 The above equations can now be solved by making the ratio exact differential,
(following from eq.(B5) of appendix B)
\begin{eqnarray}
{{\rm d}F_k \over G_{k}} 
 ={{\rm d}y_k \over \delta_{k1}}
\end{eqnarray}
where ${\rm d}F_k \equiv \sum_{i,j} z^{(k)}_{ij} {\rm d}b_{ij}$ and 
 $G_k \equiv \sum_{i,j} z^{(k)}_{ij} f_{ij}$ with
$z^{(k)}$ as arbitrary functions. Using eq.(C7), the $G_k$ can be shown to be 
\begin{eqnarray}
G_k={4\over (1-4x^2)^2 a^2} \left[(z^{(k)}_{11}+z^{(k)}_{33})[(\gamma a -4)-4x^2(\gamma a+1)]
+(z^{(k)}_{13}+z^{(k)}_{31}) 4x (8-\gamma a (1-4x^2)) +
+ 16 z^{(k)}_{22} a^{-2} x_1 (\gamma a-16 x_1)  
- z^{(k)}_{44} a x_1^{-1} (2\gamma x_1 + a)/2 \right]
\end{eqnarray}
Similarly
\begin{eqnarray}
{\rm d}F_k \equiv {\sum_{ij} z^{(k)}_{ij} {\rm d}b_{ij} 
= \left[(z^{(k)}_{11}+z^{(k)}_{33}) {\rm d}x_2 -2 (z^{(k)}_{13}+z^{(k)}_{31}) {\rm d}(xx_2) 
+ 2 (4 z^{(k)}_{22}-z^{(k)}_{44}) {\rm d} (x_1/a) \right]   
}\end{eqnarray}
where $x_2=a^{-1}(1-4x^2)^{-1}$. 
As both ${\rm d}F_1$ and $G_1$ are functions of $l_c$ and $\sigma$ (through $a$ and $\alpha$), 
 $y_1$ will turn out to be a function of parameter $l_c$ and $\sigma$,  
\begin{eqnarray}
y_1 =  \sum_{ij} \int {\rm d}b_{ij} {z^{(1)}_{ij} \over G_1} = 
\int {{\rm d}F_1(l_c,\sigma)\over G_1(l_c,\sigma)} 
\end{eqnarray}

Proceeding similarly for $y_k$ $k>2$, a solution for $y_k$ can be given as 
\begin{eqnarray}
y_k =  \sum_{ij} \int {\rm d}b_{ij} z^{(k)}_{ij} 
\end{eqnarray}
where $z^{(k)}$ satisfy the conditions $G_k=0$ 

Note the condition $G_2=0$ is satisfied for a following choice of $z^{(2)}$: 
$z^{(2)}_{11}=-z^{(2)}_{33}$, $z^{(2)}_{13}=-z^{(2)}_{31}$ and $z^{(2)}_{22}=z^{(2)}_{44}=0$. 
The $y_2$ for this choice turns out to be a constant.  
Similarly the condition $G_3=0$ can be satisfied for a following choice of $z^{(3)}$: 
${z^{(3)}_{11}+z^{(3)}_{33}\over z^{(3)}_{13}-z^{(3)}_{31}}= 
{4x{8-\gamma a(1-4x^2)}\over{(\gamma a -4)-4(\gamma a+1)x^2}}$ and 
${z^{(3)}_{22}\over z_{44}} = {a^3 x_1^3 (\gamma+a x_1/32)\over 64 (\gamma a x_1-16)}$. 
Using these $z$ values in eq.(c13), one can obtain $y_3$ as a function of $\sigma$ and $l_c$.
Note although $y_3$ varies with changing $\sigma$ and $l_c$, however 
$\sum_{ij} f_{ij}{\partial y_3\over \partial b_{ij}} =0$ and therefore $y_3$ does not 
affect  the evolution of $\rho(H)$.    
   
\end{appendix}

%\end{multicols}

\end{document}